\documentclass[usenatbib]{mn2e}
\usepackage{epsfig}
\usepackage{amsmath}
\usepackage{ulem}

\def\gsim{\mathrel{\raise0.35ex\hbox{$\scriptstyle >$}\kern-0.6em 
\lower0.40ex\hbox{{$\scriptstyle \sim$}}}}
\def\lsim{\mathrel{\raise0.35ex\hbox{$\scriptstyle <$}\kern-0.6em 
\lower0.40ex\hbox{{$\scriptstyle \sim$}}}}
\def\gs{\mathrel{\raise0.35ex\hbox{$\scriptstyle >$}\kern-0.6em 
\lower0.40ex\hbox{{$\scriptstyle \sim$}}}}
\def\ls{\mathrel{\raise0.35ex\hbox{$\scriptstyle <$}\kern-0.6em 
\lower0.40ex\hbox{{$\scriptstyle \sim$}}}}

\def\oii{{\rm [O{\sc ii}]}}

\def\kms {{\,\rm km\,s^{-1}}}

\def\lesssim{\mathrel{\hbox{\rlap{\hbox{\lower4pt\hbox{$\sim$}}}\hbox{$<$}}}}
\def\gtrsim{\mathrel{\hbox{\rlap{\hbox{\lower4pt\hbox{$\sim$}}}\hbox{$>$}}}}

\date{\today}
\title[Large-Scale Structure at $z=0.55$]
{A Huge Filamentary Structure at $z=0.55$ and Star Formation Histories of Galaxies at $z<1$}

\author[Tanaka et al.]{
\parbox[t]{\textwidth}{
Masayuki Tanaka$^{1,2}$,
Takako Hoshi$^3$,
Tadayuki Kodama$^4$,
Nobunari Kashikawa$^4$
}
\vspace*{6pt}\\
$^{1}$Department of Astronomy, School of Science, University of Tokyo, Tokyo 113--0033, Japan \\
$^{2}$European Southern Observatory, Karl-Schwarzschild-Str. 2 D-85748 Garching bei M\"{u}nchen, Germany\\
$^{3}$Department of Physics, Meisei University, 2-1-1 Hodokubo, Hino, Tokyo 191--8605, Japan\\
$^{4}$National Astronomical Observatory of Japan, Mitaka, Tokyo 181--8588, Japan \\
}

\begin{document}

\maketitle

\begin{abstract}
We report a definitive confirmation of a large-scale structure
around the super rich cluster CL0016$+$1609 at $z=0.55$.
We made spectroscopic follow-up observations with FOCAS on Subaru
along the large filamentary structure identified in our previous
photometric studies, including some subclumps already found by other
authors.
We have confirmed the physical connection of the huge filament
extending over 20$h_{70}^{-1}$ Mpc in the N-S direction hosting the
main cluster and several clumps aligned in a chain-like structure.
We have also confirmed a physical association of a new filament
extending from the main cluster to the East, which was newly discovered by us.
Based on a simple energy argument, we show that it is likely that
most of the clumps are bound to the main CL0016 cluster.
Given its spatial extent and richness, this structure is surely
one of the the most prominent confirmed structures ever identified
in the distant Universe, which then serves as an ideal laboratory to
examine the environmental variation of galaxy properties.
We draw star formation histories of galaxies from the composite spectra
of red galaxies in field, group, and cluster environments.
Combining the results from our previous studies, we find that
red galaxies in groups at $z\sim0.8$ and red field galaxies at $z\sim0.5$
show strong H$\delta$ absorption lines for their $D_{4000}$ indices.
These are the environments in which we observed the on-going build-up
of the colour-magnitude relation in our previous photometric analyses.
The strong H$\delta$ absorption lines observed in their composite spectra
imply that their star formation is truncated on a relatively short time scale
in these relatively low density environments.
We suggest that a galaxy-galaxy interaction is the most likely
physical driver of the truncation of star formation and
thus responsible for the build-up of the colour-magnitude
relation since $z\sim1$.
\end{abstract}


%
%
\section{Introduction}
\label{sec:intro}

Galaxy formation takes place in high density peaks of the density fluctuations of the Universe.
This spatially inhomogeneous galaxy formation results in large-scale structures in the Universe.
Structures develop with time and galaxies live in fine filamentary structures
in the local Universe as probed by various redshift surveys
such as the Sloan Digital Sky Survey \citep{york00}.
We often see massive concentrations of galaxies at intersections between the filaments.
Numerous studies of galaxy properties (see \citealt{tanaka05} and references therein)
suggest that galaxies in different places in the Universe live a totally different life
--- galaxies in clusters tend to end up with red early-type galaxies, while
those in the field tend to be blue late-type galaxies at the present day.
Environment appears to play an important role in determining the fate of a galaxy.

To quantify the galaxy evolution as a function of environment and to get
a handle on the origin of the environmental dependence of galaxy properties,
we are conducting a high redshift cluster survey PISCES with the Subaru telescope \citep{kodama05}.
We map out large-scale structures surrounding high redshift clusters
with the unique wide-field imager Suprime-Cam \citep{miyazaki02}.
We then perform intensive spectroscopic follow-up observations
to examine the structure evolution and star formation histories of galaxies.
Recent results from our survey are reported in \citet{kodama05}, \citet{nakata05},
and \citet{tanaka05,tanaka06,tanaka07}.

In this paper, we focus on one of the PISCES clusters, CL0016$+$1609 at $z=0.55$.
It is one of the most extensively studied galaxy clusters
\citep{butcher84,ellis97,dressler99,brown00,worrall03,zemcov03,dahlen04}.
As reported in \citet{kodama05}, we observed the cluster in $BVRi'z'$ with
Suprime-Cam under good sky conditions.
We applied the photometric redshift technique to extract galaxies at the cluster
redshift and discovered a prominent large-scale structure around the cluster.
The cluster accompanies several clumps around it
(see also \citealt{koo81,hughes95,connolly96,hughes98}).
These clumps and the cluster appear to be embedded in a huge filament extending over 20Mpc.

A cautionary note here is that photometric identification of structure
is subject to projection effects.
Also, the limited accuracy of photometric redshifts does not allow us
to see if the discovered clumps are located at the same redshift
and form a single structure.
They could lie at slightly different redshifts and could be dynamically independent.
Therefore, spectroscopic follow-up observations are essential to confirm
the structure.

Another important product of spectroscopic observation is that spectra of galaxies
provide us with information on star formation histories of galaxies, which cannot be
easily inferred from photometry.
\citet{tanaka05} presented the build-up of the colour-magnitude relation.
This build-up involves a suppression of star formation activities in galaxies.
What physical process(es) suppresses star formation and makes blue galaxies red?
We can put a constraint on the physical drivers of the suppression of
star formation with spectroscopic information.

We have made a spectroscopic observation of CL0016
and we report on the results in this paper.
We spectroscopically confirm a huge filamentary structure extending
N-S direction over 20 Mpc from the cluster and another filament in the E-W direction.
These form one of the the most prominent structures at high redshifts discovered so far.
We also focus on star formation histories of galaxies in the structures.
In particular, we discuss star formation histories of red galaxies
in parallel to the build-up of the colour-magnitude relation.

The layout of this paper is as follows.
We describe our observation in Section \ref{sec:obs} and
present the spectroscopically confirmed large-scale structure in Section \ref{sec:lss}.
We then turn our attention to star formation histories of galaxies
in the structure in Section \ref{sec:sfh}.
Section \ref{sec:discussion} discusses physical drivers of star formation truncation
and the build-up of the colour-magnitude relation.
Finally, the paper is summarized in Section \ref{sec:summary}.
Our analysis scheme closely follows that adopted in \citet{tanaka06}.
We refer the reader to the paper for details of the scheme.

Throughout this paper, we assume a flat Universe with
$\Omega_{\rm M}=0.3,\ \Omega_{\rm \Lambda}=0.7$ and $H_0=70\kms \rm Mpc^{-1}$.
Magnitudes are on the AB system.
We use the following abbreviations:
CMD for colour-magnitude diagram and CMR for colour-magnitude relation.

%
%
\section{Observation and Data Reduction}
\label{sec:obs}

We conducted spectroscopic follow-up observations of CL0016
during 11--14 October 2004 and 19--20 January 2007 with FOCAS
\citep{kashikawa02} in MOS mode.
The instrumental configuration and observation strategy were
the same as reported in \citet{tanaka06}.
We used a 300 lines $\rm mm^{-1}$ grating
blazed at 5500 $\rm \AA$ with the order-cut filter SY47.
The wavelength coverage was between 4700$\rm\AA$ and 9400$\rm\AA$ with a pixel
resolution of $1.4\rm \AA\  pixel^{-1}$.
A slit width was set to $0\farcs 8$, which gave a resolution of
$\lambda/\Delta\lambda\sim500$.

\begin{figure*}
\begin{center}
\leavevmode
\epsfxsize 0.6\hsize \epsfbox{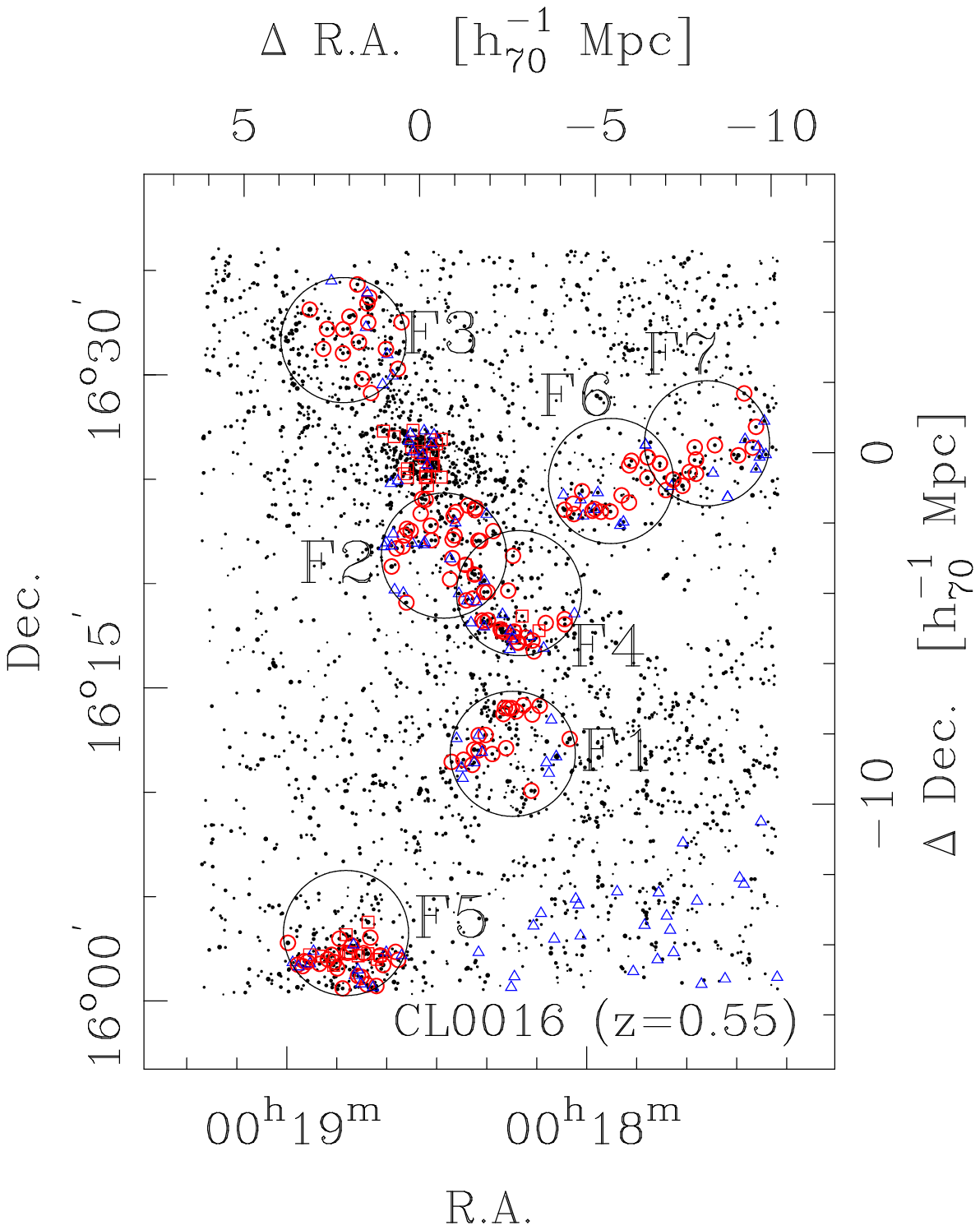}
\end{center}
\caption{
Spectroscopic target fields (F1--F7) in CL0016 as shown by the large circles.
The points show photo-$z$ selected galaxies at $0.48\leq z_{phot}\leq0.60$.
The small open circles are our spectroscopic objects at $0.53<z_{spec}<0.56$
with secure redshift estimates ($z_{conf}=0$).
The squares are spectroscopic objects drawn from  \citet{hughes95},
\citet{munn97}, \citet{hughes98}, and \citet{dressler99} in the same redshift interval.
The triangles show galaxies outside the interval including both our sample
and those from the literature.
The top and right ticks show the comoving scales.
North is up and East is to the left.
}
\label{fig:target_fields}
\end{figure*}

\begin{figure}
\begin{center}
\leavevmode
\epsfxsize 0.8\hsize \epsfbox{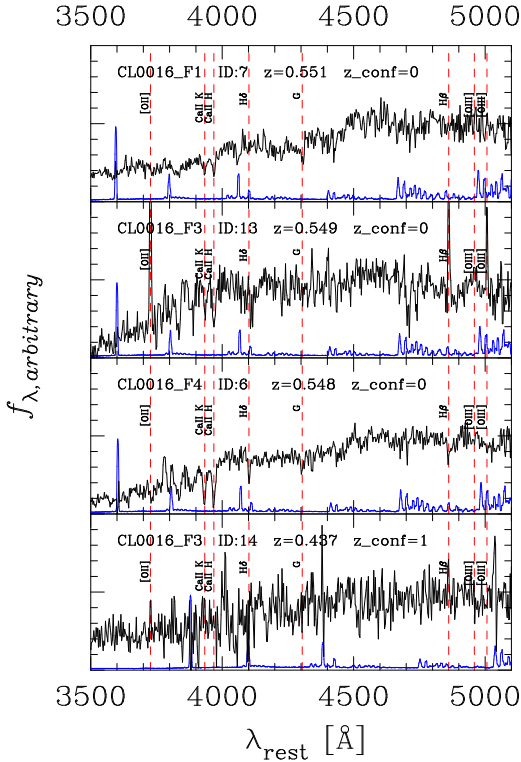}
\end{center}
\caption{
Spectra obtained with FOCAS.  The sky emission lines are drawn at the bottom in each panel.
}
\label{fig:spec_example}
\end{figure}

Data reduction and analyses scheme closely follows that adopted in \citet{tanaka06}.
Here we describe only the outline of it.
We selected 7 FOCAS fields which efficiently cover the photometrically identified
large-scale structure as shown in Fig. \ref{fig:target_fields}.
Bright galaxies ($m_{i'}\lesssim22$) at $0.48\leq z_{phot}\leq 0.60$
were given the highest priority in the slit assignment.
About 74  per cent of the targeted galaxies (185 out of 251)
fall in this photo-$z$ range.
Total on-source exposures are listed in Table \ref{tab:obs_summary}.
Data reduction is performed in a standard manner using {\sc IRAF}.
All the reduced spectra are visually inspected using custom designed
software and redshift estimates and confidence flags are assigned.
We obtain 195 secure redshifts out of 251 observed galaxies.
We present in Fig. \ref{fig:spec_example} some of our spectra
in the CL0016 field.
Table \ref{tab:spec_catalog} gives a catalogue of our spectroscopic objects.

To assess the accuracy of the spectroscopic redshifts,
we have cross-correlated our spectroscopic objects with those from
\citet{hughes95} and \citet{hughes98}.
Six objects are matched.
Redshifts of five objects agree within the errors, and the median of
$z_{spec,hughes95,98}-z_{spec,tanaka}$ is 0.00001
and the dispersion around it is 0.0004.
However, one object has a deviant redshift
$z_{spec,hughes95}=0.42000$, while $z_{spec,tanaka}=0.08160$.
The redshift estimated by \citet{hughes95} is of low confidence (see their Table 2).
But, this object (Field ID: F4, Slit ID:16) has a confident redshift
in our catalogue and photo-$z$ is consistent with our measurement ($z_{phot}=0.05$).
Thus, we take our measurement.
We also check internal consistency using objects with multiple observations.
Four objects are observed in both F6 and F7 (F6-31 and F7-1, F6-32 and F7-2,
F6-34 and F7-4, and F6-36 and F7-5).
The median of the difference between the two measurements is 0.00007 and
dispersion around it is 0.00069, revealing a good consistency.
In what follows, we use galaxies with secure redshifts only to avoid
any possible uncertainties and biases.
Finally, our spectroscopic catalogue is combined with those from \citet{hughes95},
\citet{munn97}, \citet{hughes98}, and \citet{dressler99}
\footnote{
\citet{munn97} and \citet{dressler99} assigned their own confidence flags ($q$)
to their redshifts.  Here we adopt $q\geq3$ in \citet{munn97}
and $q\leq3$ in \citet{dressler99} as secure redshifts.
}.
This makes a large spectroscopic sample of 281 galaxies in and around the cluster.

\begin{table*}
\caption{
Log of the spectroscopic observation.
}
\label{tab:obs_summary}
\begin{tabular}{ccccl}\hline
Field ID & Date       &  R.A. (J2000)           & Dec. (J2000)             & Exposures \\
\hline
F1      & 2004-10-11/12 &  $00^h\ 18^m\ 14^s.90$  & $+16^\circ\ 11'\ 50''.0$ &  1500s $\times$ 3shots\\
F2      & 2004-10-12 &  $00^h\ 18^m\ 28^s.60$  & $+16^\circ\ 21'\ 20''.0$ &  1200s $\times$ 4shots\\
F3      & 2004-10-13 &  $00^h\ 18^m\ 48^s.66$  & $+16^\circ\ 31'\ 39''.8$ &  1500s $\times$ 3shots\\
F4      & 2004-10-14 &  $00^h\ 18^m\ 13^s.60$  & $+16^\circ\ 19'\ 32''.0$ &  1500s $\times$ 3shots\\
F5      & 2004-10-14 &  $00^h\ 18^m\ 48^s.20$  & $+16^\circ\ 03'\ 14''.8$ &  1500s $\times$ 3shots\\
F6      & 2007-01-19 &  $00^h\ 17^m\ 55^s.20$  & $+16^\circ\ 24'\ 54''.0$ &  1200s $\times$ 3shots\\
F7      & 2007-01-20 &  $00^h\ 17^m\ 36^s.00$  & $+16^\circ\ 26'\ 42''.0$ &  1200s $\times$ 3shots\\
\hline
\end{tabular}
\end{table*}

\begin{table*}
\caption{
Catalogue of our spectroscopic objects.
{\it This table will appear in its entirety in the electric edition of the journal.}
}
\label{tab:spec_catalog}
\begin{tabular}{llllrrrrllllll}
\hline
ID & R.A. & Dec. & $m_{z', tot}$ & $B-V$ & $V-R$ & $R-i'$ & $i'-z'$ &
$z_{phot}$ & $z_{spec}$ & $z_{spec, min}$ & $z_{spec, max}$ & $z_{conf}$\\
\hline
F1-1 & 00 18 9.52 & 16 14 7.8 & 20.06 & 1.11 & 1.12 & 0.63 & 0.45 & 0.54 & 0.5497 & 0.5495 & 0.5499 & 0\\
F1-2 & 00 18 16.47 & 16 14 2.2 & 19.10 & 1.61 & 1.32 & 0.78 & 0.47 & 0.55 & 0.5478 & 0.5475 & 0.5480 & 0\\
F1-3 & 00 18 15.76 & 16 14 0.6 & 22.06 & 1.51 & 1.32 & 0.69 & 0.42 & 0.52 & 0.5437 & 0.5433 & 0.5441 & 0\\
F1-4 & 00 18 14.18 & 16 13 52.1 & 19.81 & 1.51 & 1.28 & 0.77 & 0.48 & 0.55 & 0.5504 & 0.5497 & 0.5509 & 0\\
F1-5 & 00 18 12.74 & 16 14 9.2 & 21.24 & 0.93 & 1.04 & 0.59 & 0.43 & 0.56 & 0.5497 & 0.5492 & 0.5501 & 0\\
F1-6 & 00 18 10.99 & 16 13 42.9 & 22.01 & 0.86 & 0.88 & 0.41 & 0.33 & 0.47 & 0.5428 & 0.5426 & 0.5430 & 0\\
F1-7 & 00 18 16.75 & 16 13 44.5 & 20.90 & 1.56 & 1.27 & 0.73 & 0.46 & 0.55 & 0.5512 & 0.5506 & 0.5514 & 0\\
F1-8 & 00 18 17.40 & 16 13 36.1 & 21.92 & 1.37 & 1.18 & 0.67 & 0.41 & 0.54 & 0.5469 & 0.5461 & 0.5475 & 1\\
F1-9 & 00 18 11.19 & 16 10 3.8 & 21.32 & 0.64 & 0.79 & 0.31 & 0.21 & 0.50 & 0.5539 & 0.5538 & 0.5540 & 0\\
F1-10 & 00 18 20.47 & 16 14 3.1 & 19.61 & 1.26 & 1.22 & 0.71 & 0.47 & 0.55 & 0.5513 & 0.5503 & 0.5521 & 1\\
F1-12 & 00 18 20.30 & 16 12 44.0 & 20.46 & 1.55 & 1.28 & 0.73 & 0.46 & 0.54 & 0.5486 & 0.5481 & 0.5489 & 0\\
F1-13a & 00 18 21.69 & 16 12 43.6 & 21.86 & 0.71 & 0.74 & 0.32 & 0.15 & 0.47 & 0.4813 & 0.4811 & 0.4817 & 0\\
F1-13b & 00 18 21.69 & 16 12 43.6 & 21.86 & 0.71 & 0.74 & 0.32 & 0.15 & 0.47 & 0.5529 & 0.5526 & 0.5532 & 0\\
F1-14 & 00 18 21.75 & 16 11 59.7 & 20.70 & 1.22 & 1.23 & 0.68 & 0.50 & 0.55 & 0.5518 & 0.5516 & 0.5520 & 0\\
F1-15 & 00 18 22.60 & 16 12 2.0 & 19.95 & 1.43 & 1.29 & 0.74 & 0.51 & 0.55 & 0.5540 & 0.5539 & 0.5541 & 0\\
\hline
\end{tabular}
\begin{flushleft}
The format of ID is ``Field ID -- Slit ID''.
The astrometric calibration is performed against the USNO2-B catalogue \citep{monet03}.
An accuracy of our coordinates is $\sim0''.2$.
Total magnitudes ($m_{z',tot}$) are measured in Kron-type apertures ({\sc MAG\_AUTO}),
while colours are measured within $2''$ apertures.
Objects with secure/probable redshift estimates are flagged as $z_{conf}=0$
and those with likely redshifts as $z_{conf}=1$.
Note that the redshift errors shown here do not include the error in the wavelength calibration,
which is typically $\sim0.3\rm\AA$.
In some cases, two redshifts are obtained for the same object (e.g., F1-13).
It is likely that two objects lie on the line-of-sight.
F5-28 has no photometric redshift due to its close proximity to a bright star.
\end{flushleft}
\end{table*}

%
%
\section{Large-Scale Structure at $z=0.55$}
\label{sec:lss}

\subsection{Discovery of Large-Scale Structure}
We briefly review the photometrically identified large-scale structure
around CL0016 (see Fig. 5 of \citealt{kodama05}).
A very rich cluster lies at NE from the centre of the field.
This rich cluster is known to accompany two clumps
seen at ($\rm\Delta R.A.,\ \Delta Dec.$)=($-4',\ -8'$) and ($+4',\ -24'$)
relative to the main CL0016 cluster \citep{hughes95,connolly96,hughes98}.
We discover a new clump at ($-4',\ -13'$).
A filament appears to connect all these
clumps and forms a huge ($>20\rm\ h_{70}\ Mpc^{-1}$ comoving)
structure extending in the N-S direction.
A few more filaments going in the E-W direction can also be seen
[e.g. starting from ($-8',\ -3'$) and extends westward].
All these well-visible structures form one of the most prominent
structures at high redshifts discovered so far.
We have confirmed that almost all of the structures can be reproduced
even if we apply colour cuts to extract only the red
cluster member candidates along the colour-magnitude sequence
instead of applying the photo-$z$ selection \citep{kodama05}.
The observed structure is therefore unlikely to be the artifacts of
the photometric redshifts.

\begin{figure*}
\begin{center}
\leavevmode
\epsfxsize 0.42\hsize \epsfbox{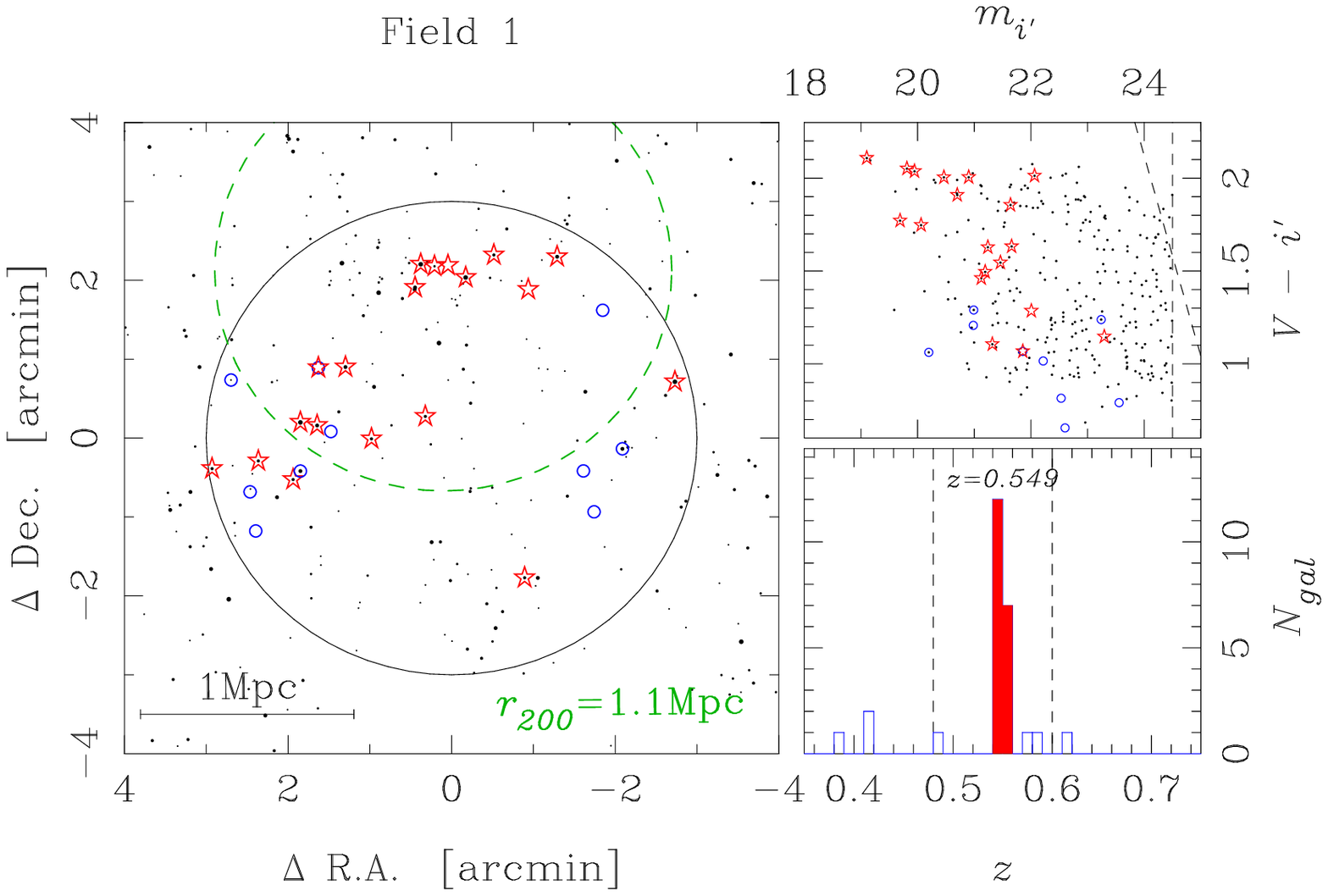}\hspace{0.5cm}
\epsfxsize 0.42\hsize \epsfbox{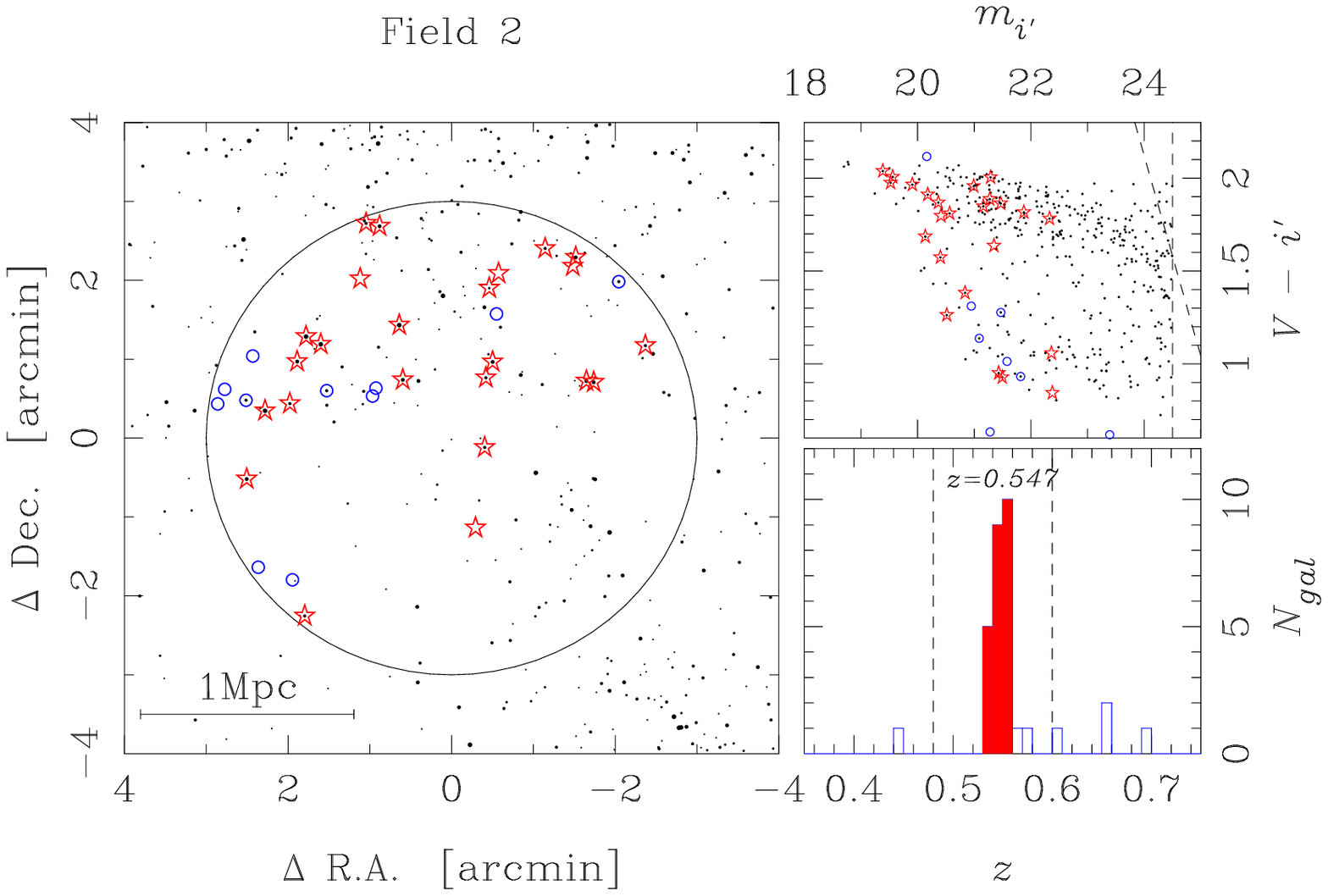}\\\vspace{0.3cm}
\epsfxsize 0.42\hsize \epsfbox{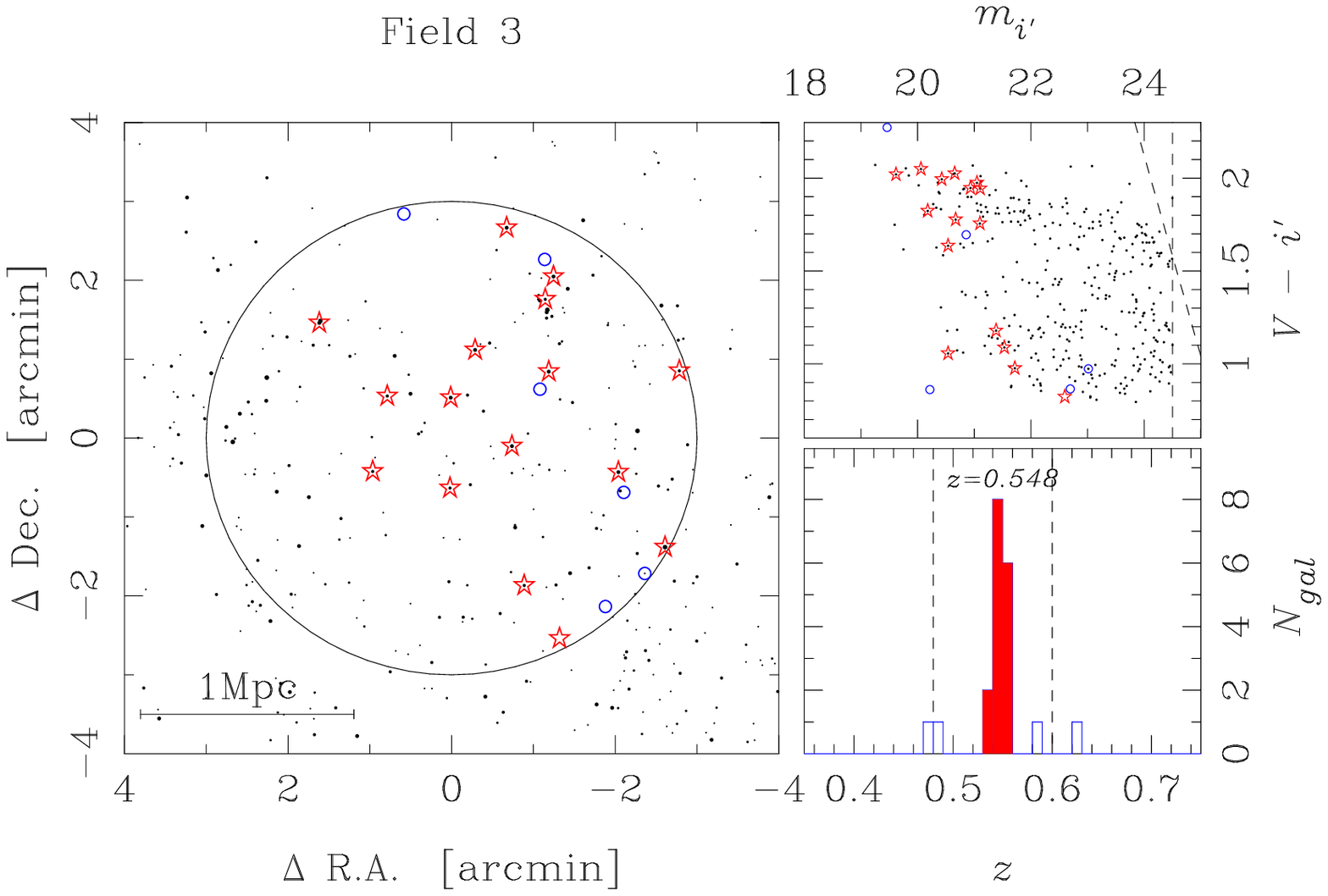}\hspace{0.5cm}
\epsfxsize 0.42\hsize \epsfbox{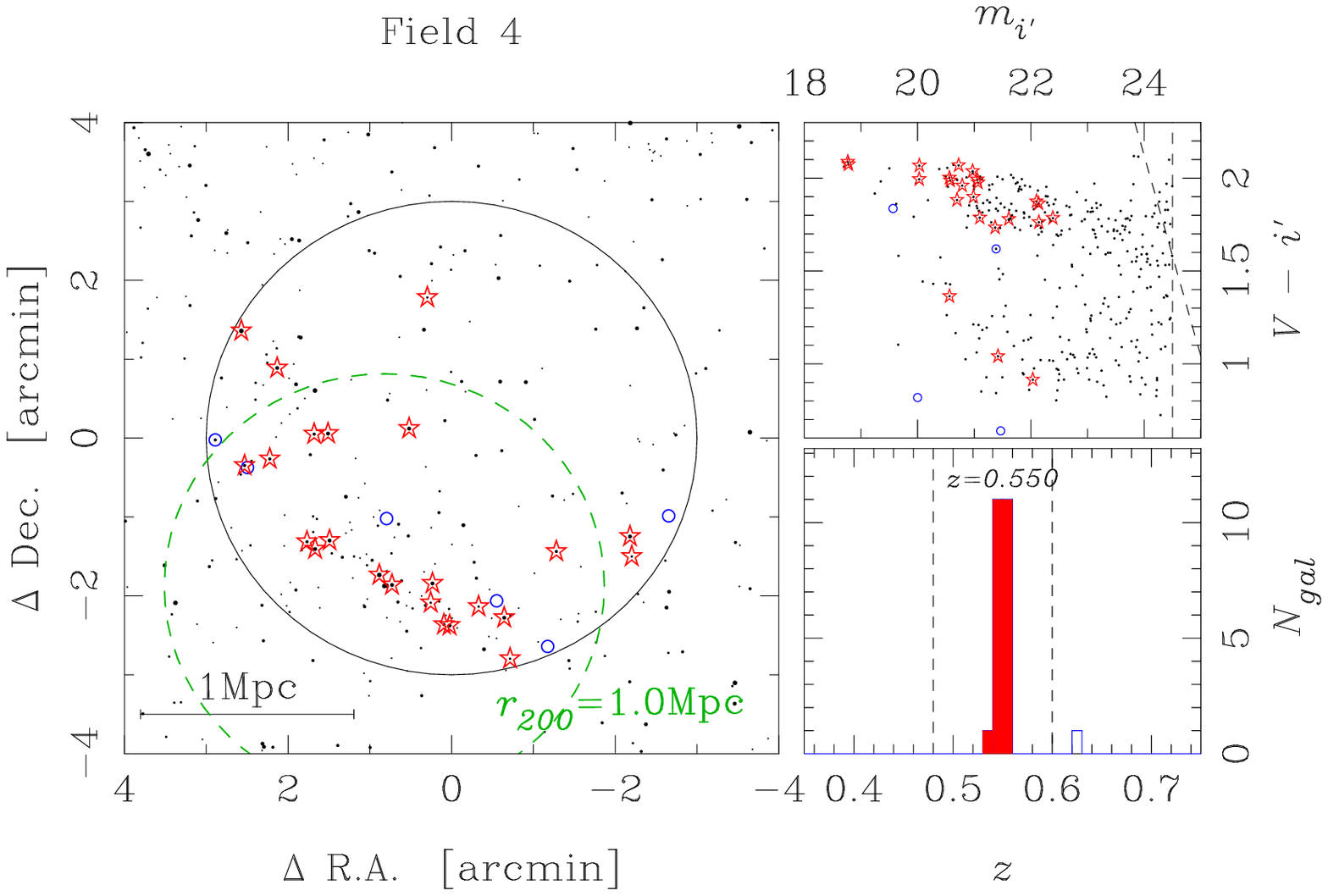}\\\vspace{0.3cm}
\epsfxsize 0.42\hsize \epsfbox{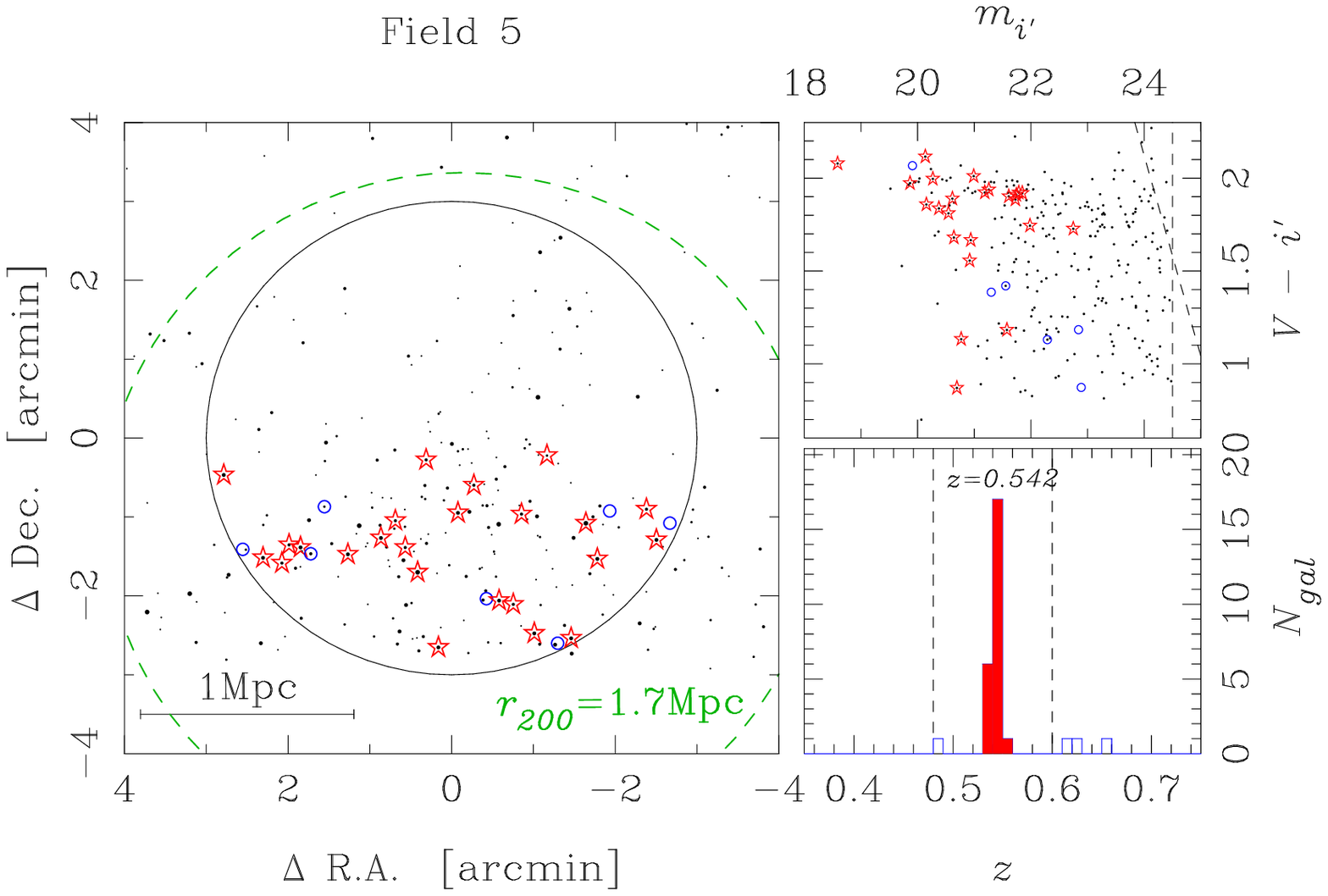}\hspace{0.5cm}
\epsfxsize 0.42\hsize \epsfbox{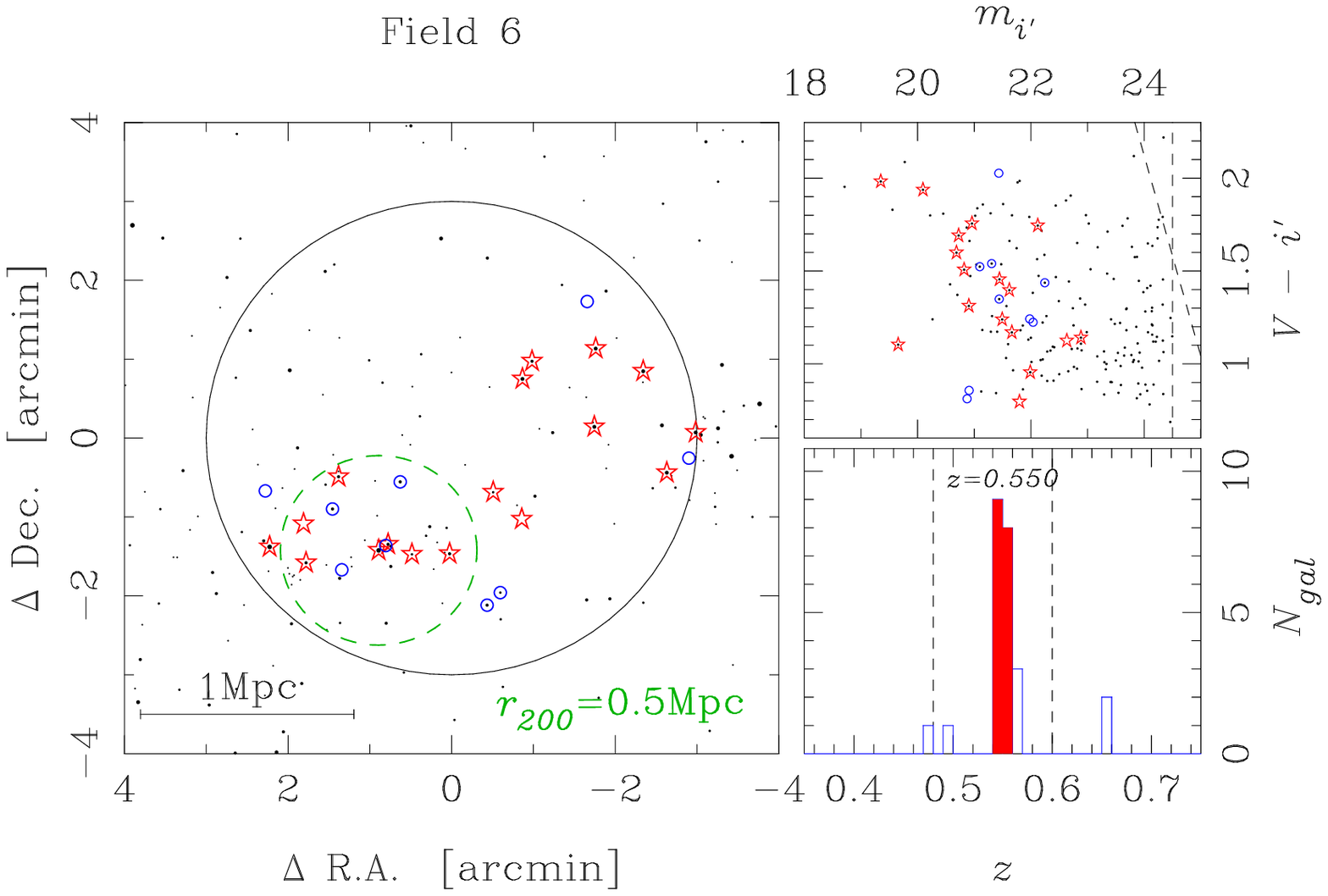}\\\vspace{0.3cm}
\epsfxsize 0.42\hsize \epsfbox{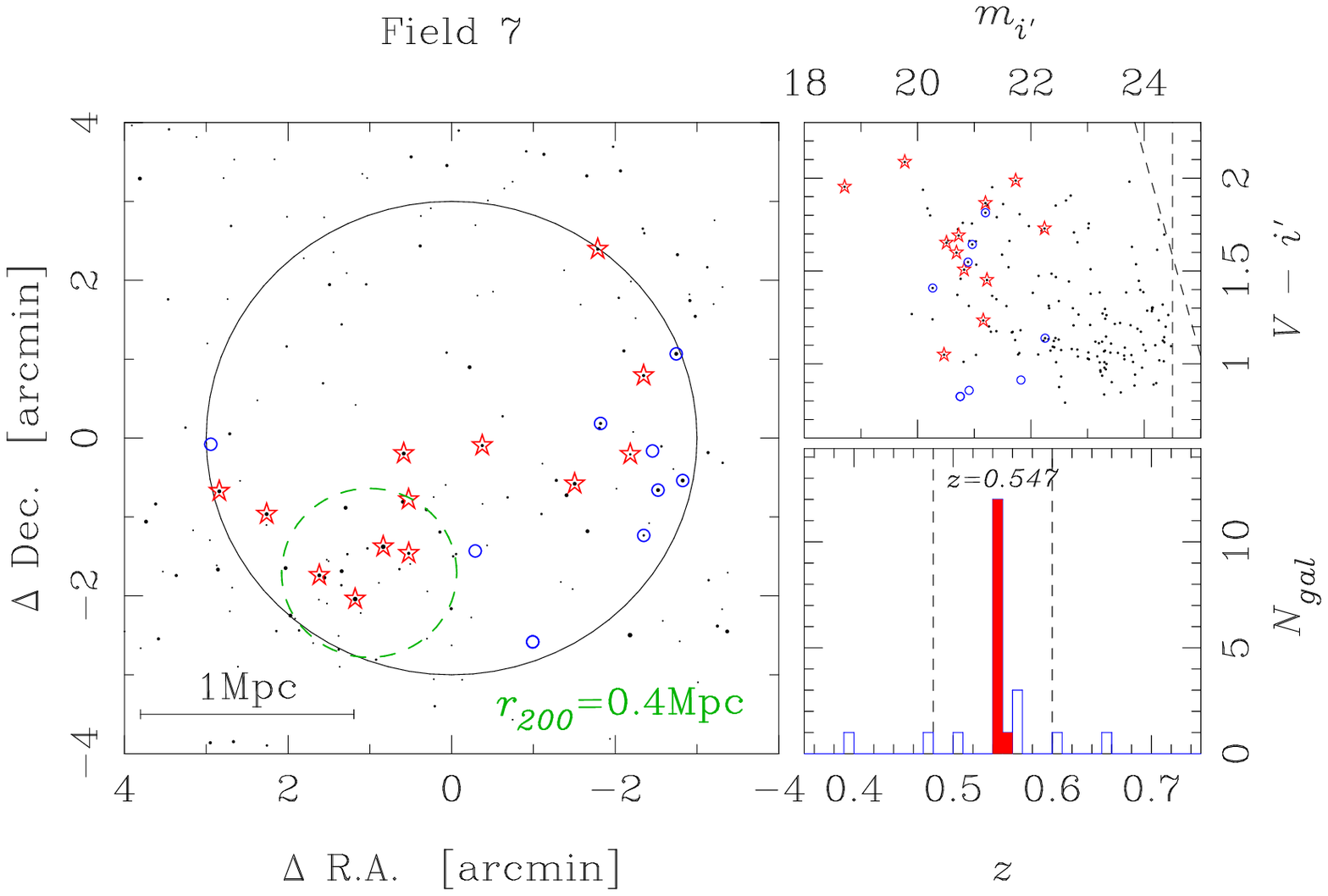}
\end{center}
\caption{
Close-up views of the target fields in CL0016.
{\it Bottom-right panel in each plot:}
The redshift distribution of spectroscopically observed galaxies in each field.
The filled histograms show a redshift spike at $0.53<z<0.56$ and
galaxies in this spike are shown as stars in the other panels.
The vertical dashed lines mean our primary photo-$z$ selection range.
{\it Top-right panel in each plot:}
The $V-i'$ colour plotted against the $i'$-band magnitude
using galaxies at $0.48\leq z_{phot}\leq0.60$.
The stars are the spectroscopic objects in the redshift spike,
and the open circles are objects outside of this spike.
The dashed lines show the magnitude cut and $5\sigma$ limiting colours.
{\it Left panel in each plot:}
The points show the distribution of photo-$z$ selected
($0.48\leq z_{phot}\leq0.60$) galaxies, and the circle shows
the field of view of FOCAS.
The 1-Mpc scale is expressed as the physical distance.
The stars/circles are galaxies inside/outside the redshift spike, respectively.
The virial radius of a system is plotted as the dashed circle in some plots.
}
\label{fig:close_up_views}
\end{figure*}

We summarize results of the spectroscopic observations in Fig. \ref{fig:close_up_views}.
Note that only our spectroscopic objects are shown in the figure, and those
from the literature are not plotted.
Interestingly, spectroscopic galaxies in each field show a sharp redshift spike
around the cluster redshift of $z=0.5484\pm0.0014$, which is estimated
from the spectroscopic data from \citet{dressler99} (we use galaxies at $0.535<z<0.558$,
see discussions below).
The redshift spikes are much narrower than the photo-$z$ selection range,
which is indicated by the vertical dashed lines.
Therefore, the spikes are not products of selection effects but are real structures in each field.
Table \ref{tab:vel_disp} presents the redshift centres of the redshift spikes.
All the redshift spikes are located very close to the redshift of the main cluster, $z=0.548$.
The clump in F5 lies at slightly lower redshift
($\Delta z=-0.006$ or $\sim20\rm h_{70}^{-1}$ Mpc in comoving scale).
The structure might bend along the line-of-sight at South.
Spectroscopically confirmed structure members ($0.53<z<0.56$) are shown
in  Fig. \ref{fig:target_fields} as the open circles and squares.
Spectroscopic objects from the literature are included in this plot.
It is now clear that a huge filament goes in the N-S direction
and another filament extends from the main cluster towards west.
We note that 175 out of the 281 galaxies with secure redshifts lie at
$0.53 < z_{spec} < 0.56$.

Galaxies in F1, 4, 5, 6 and 7 are clearly clustered and we estimate
their centres and virial radii as follows.
In F4, 5 and 6, there is an outstandingly bright galaxy, which is
a spectroscopically confirmed member, and we take the position of
the galaxy as the centre.
There is no such bright galaxy in F1 and 7.
We take the average of a few brightest spectroscopically confirmed
members as the centre.
We adopt $r_{200}$ as the virial radius \citep{carlberg97},
and we apply an iterative estimate of $r_{200}$.
First, we estimate $r_{200}$ using all the spectroscopic members
in each field.  Then, we re-estimate $r_{200}$ using
galaxies within the firstly estimated $r_{200}$.
We use the $2\sigma$-clipped biweight and gapper estimates
for the central redshift and velocity dispersion
(which is then translated into $r_{200}$), respectively \citep{beers90}.
The estimated $r_{200}$ is shown in Fig. \ref{fig:close_up_views}.

This strong clustering suggests that these are gravitationally bound systems.
In contrast, galaxies in F2 and F3 show no clear spatial concentrations.
F2 is likely the outskirts of the cluster 
connecting the main cluster and the clump in F4.
F3 has somewhat large velocity dispersion ($\sigma=930\kms$) 
and it is unlikely a bound system.
But, F3 covers some compact groups seen in the distribution of photo-$z$ selected galaxies
(see Fig. 6 of \citealt{tanaka05}), which we may miss in the spectroscopic sample
due to the poor spatial sampling.
Also, the loose galaxy distribution in F3 (i.e.  no clear concentrations)
may suggest that this region contains galaxies in the adjacent filament that connects
the clumps.
We note that F3 provides us with evidence that the structure extends
northward of the main cluster.

Following the confirmation of the filamentary and clumpy structures,
we make a simple energy argument to evaluate the probabilities
that the clumps are bound to the main CL0016 cluster
in the next subsection.

\subsection{Dynamical Analysis}

We derive the probabilities that the clumps and the CL0016 main cluster
are bound systems based on the Newtonian energy integral formalism
\citep{beers82,hughes95,lubin98}.
We introduce a positional angle $\phi_d$ between a clump and
the main cluster relative to the line-of-sight.
The linear distance between them is then given by
$d=D_p/\sin\phi_d$, where $D_p$ is the projected distance.
Similarly, the relative space velocity between them is represented as
$v=v_r/\cos\phi_v$, where $v_r$ is the line-of-sight velocity difference
and $\phi_v$ is the angle in the velocity space relative to the line-of-sight.
The condition that a clump and the main cluster is bound
is then expressed by

\begin{equation}\label{eq:bound_prob}
v_r^2-(2GM/D_p)\sin\phi_d\cos^2\phi_v<0.
\end{equation}

\noindent
where $M$ is the mass of the system.
The probability that the group is dynamically bound to the main cluster is
estimated by integrating the solid angle ($\phi_d$ and $\phi_v$)
that satisfies the condition.

We summarize in Table \ref{tab:system_props} properties of each clump.
The mass within $r_{200}$ (denoted as $M_{200}$) is evaluated following
the prescription in \citet{carlberg97}.
The redshift and mass of the main cluster are estimated to be
$z=0.5484\pm0.0014$ and $M_{200}=3.0^{+1.4}_{-1.0}\times10^{15}$
using galaxies at $0.535<z<0.558$ from \citet{dressler99}.
We refer to these values in the following analysis, but
the cluster galaxies show a redshift tail on both higher and lower
redshifts.
If we adopt a wider range of $0.520<z<0.570$, we obtain
$z=0.5472\pm0.0017$ and $M_{200}=5.9^{+3.2}_{-2.3}\times10^{15}$,
but our conclusion below remain unchanged.

Putting all these values into Eq. \ref{eq:bound_prob},
we find that it is very likely that the clumps are dynamically
bound to the main cluster.  The probability that the clump in F1
is bound is $88^{+8}_{-17}$\% (the error is based on a Monte-Carlo simulation).
The bound probabilities for the clumps in F4, 5, 6, and 7 are $79^{+14}_{-22}$\%,
$26^{+22}_{-19}$\%, $69^{+13}_{-15}$\%, and  $90^{+6}_{-16}$\%, respectively.
The probability for the clump F5 is small compared with other clumps,
but this is due to its slightly low redshift from the main cluster
and somewhat large projected distance.
Note that \citet{hughes95} obtained a 38 per cent probability that
this clump is bound.
The difference from ours is due to the difference in the spectroscopic
sample (we have more spec-$z$) and the different technique adopted for
estimating cluster mass (\citealt{hughes95} relied on X-ray).
The probabilities given above should be considered as the lower limits since
the cluster-cluster peculiar velocity will be less than a few thousand $\kms$,
and the range of $\phi_v$ should be limited \citep{hughes95}.
Therefore, we suggest that the clumps in F1, 4, 6 and 7 are probably
bound to the main cluster, and the one in F5 may be bound.

To sum up, we spectroscopically confirm a huge filamentary structure at $z=0.55$
hosting the rich cluster and several clumps.
The clumps are likely bound to the central rich cluster, adding
further evidence for the large-scale structure at $z=0.55$.
This is one of the most prominent structure ever discovered at high redshifts.
The large scale structure provide us with galaxies in a wide range of environment,
and it serves as an ideal site to study the effects of environment on galaxy evolution.
We take this opportunity and examine star formation histories of galaxies
as a function of environment in the next section.

\begin{table*}
\caption{
Redshift of the redshift spike in each field.
$N_{member}$ is a number of spectroscopic galaxies
in the redshift spike ($0.53<z<0.56$).
The central redshifts are measured using the biweight estimator \citep{beers90}.
The errors are estimated from the jackknife resampling of the spectroscopic members.
Note that the group in F4 is identified as RXJ0018.3$+$1618 in \citet{hughes95}
and the cluster in F5 as  RXJ0018.8$+$1602 in \citet{hughes98}.
}
\label{tab:vel_disp}
\begin{tabular}{lccl}
\hline
Field ID & $N_{member}$ & $z$ & environment \\
\hline
F1  & 19 & $0.5493\pm0.0009$ & group\\
F2  & 24 & $0.5468\pm0.0016$ & cluster outskirts\\
F3  & 16 & $0.5481\pm0.0023$ & some compact groups and a possible filament\\
F4  & 23 & $0.5498\pm0.0014$ & group\\
F5  & 24 & $0.5424\pm0.0015$ & cluster\\
F6  & 17 & $0.5498\pm0.0008$ & group and filament\\
F7  & 13 & $0.5473\pm0.0004$ & group and filament\\
\hline
\end{tabular}
\end{table*}

\begin{table*}
\caption{
The dynamical properties of groups and clusters in F1, 4, 5, 6 and 7.
See text for the details of the procedure to estimate centres and
virial radii of the groups and clusters.
The redshift and velocity dispersions are measured with the biweight estimator
and gapper method, respectively \citep{beers90}.
The errors are estimated from the jackknife resampling, except these for $M_{200}$
which are from Monte-Carlo simulations.
}
\label{tab:system_props}
\begin{tabular}{cccccccc}
\hline
Field ID & R.A. & Dec. & $z$ & $\sigma\rm\ [km\ s^{-1}]$ & $D_p$ [$h_{70}^{-1}$ Mpc] & $r_{200}$ [$h_{70}^{-1}$ Mpc] & $M_{200}$ [$10^{14}\rm M_\odot$]\\ 
\hline
F1 & $00^h$ $18^m$ $15^s.3$ & $+16^\circ\ 13'\ 57''$ & $0.5496\pm0.0008$ & $583\pm185$ & 5.0 & $1.1\pm0.3$ & $2.5^{+3.2}_{-1.7}$\\
F4 & $00^h$ $18^m$ $17^s.0$ & $+16^\circ\ 17'\ 39''$ & $0.5508\pm0.0021$ & $563\pm138$ & 3.6 & $1.0\pm0.3$ & $2.3^{+2.1}_{-1.3}$\\
F5 & $00^h$ $18^m$ $47^s.6$ & $+16^\circ\ 02'\ 15''$ & $0.5422\pm0.0012$ & $903\pm136$ & 9.3 & $1.7\pm0.2$ & $9.5^{+5.0}_{-3.7}$\\
F6 & $00^h$ $17^m$ $58^s.9$ & $+16^\circ\ 23'\ 28''$ & $0.5519\pm0.0006$ & $249\pm96$  & 3.3 & $0.5\pm0.1$ & $0.20^{+0.33}_{-0.15}$\\
F7 & $00^h$ $17^m$ $40^s.2$ & $+16^\circ\ 25'\ 00''$ & $0.5474\pm0.0005$ & $221\pm69$  & 4.9 & $0.3\pm0.1$ & $0.14^{+0.17}_{-0.09}$\\
\hline
\end{tabular}
\end{table*}

%
%
\section{Star Formation Histories}
\label{sec:sfh}

Spectra of galaxies contain invaluable information on star formation histories
of galaxies, which cannot be easily inferred from broad-band photometry.
In what follows, we examine star formation histories as a function of environment.
First, we make composite spectra of red galaxies on the CMR in the cluster, group,
and field environments.
We then quantify differences in the spectra with some help of model predictions
and discuss the star formation histories in parallel to the build-up of the CMR
reported in \citet{tanaka05}.

\subsection{Composite Spectra of Red Galaxies}

We combine a number of spectra to make high-S/N average spectra
in various environments.
Here, we focus on bright red galaxies [those having $i'\lesssim21.5$  and
$\Delta(V-i')<0.15$ with respect to the CMR].
We do not examine blue galaxies because our photometric redshifts
are not accurate for blue galaxies \citep{tanaka06},
and we cannot avoid a selection bias. 
Each spectrum is normalized to unity at 4000--4200 $\rm \AA$ in
the rest frame, and a composite spectrum is then made by taking a $2\sigma$ clipped mean.
Spectra with low-$S/N$ or those affected by CCD defects are removed
when we combine the spectra.

We make a typical 'cluster' spectrum by combining spectra of red galaxies
in F5 in Fig. \ref{fig:target_fields}.
We did not observe the main cluster with FOCAS
since a lot of spectra were already taken in the MORPHS project \citep{dressler99}.
Because of this, we cannot examine the red galaxies in the core of the main cluster
within the same data set of FOCAS.
However, the clump in F5 is  rich
($\sigma\sim900\kms$; see Table \ref{tab:system_props}),
and we can define this clump as a cluster.
In fact, its velocity dispersion is comparable to that of the RXJ0153 cluster
at $z=0.83$ ($\sigma=700-900\kms$; \citealt{demarco05}), which we use later for comparison.
We recall that this clump was defined as a group in \citet{tanaka05}.
Due to the close proximity to the field edge, we fail to
correctly measure global density for this clump.
We re-analyze the CMDs at $z\sim0.5$ changing the definition
of the clump to cluster.  We confirm that conclusions in \citet{tanaka05}
are totally unchanged.

Red galaxies in F1, 4, 6 and 7 are merged into a 'group' spectrum.
Due to the prominent structure around CL0016,
we cannot find any isolated red galaxies at $0.5<z<0.6$ suitable for
a composite 'field' spectrum in our field.
Red galaxies in F2 and F3 could be used, but they are defined as
cluster/group galaxies in our previous paper \citep{tanaka05}.
In fact, F2 observes the outskirts of the cluster.
Also, small clumps of galaxies are seen in F3
in the distribution of photo-$z$ selected galaxies.
To make a fair sample of field galaxies, we collect spectra of
isolated red galaxies in CL0939 ($z=0.41$) fields.
Data are kindly provided by Nakata et al. (in prep.).
The CL0939 fields were observed with exactly the same configuration
as ours with the same spectrograph.  Thus, direct comparisons can be made.
The cluster, group, and field spectra are constructed from 11, 17, and 7
galaxies, respectively.

We quantify photometric similarities of the red galaxies.
We first correct for the small k-correction and passive evolution effects
for galaxies at slightly different redshifts \citep{kodama97}.
We then apply the 2D KS test \citep{fasano87} for the distribution of galaxies on the CMD.
It does not reject the hypothesis that the colours and magnitudes
of the cluster red galaxies and those of the field red galaxies
are drawn from the same parent population (they are from the same parent
population at 55 per cent level).
Therefore, the red galaxies in clusters and those in field are likely to
share the common photometric properties. 
The probability that the cluster red galaxies and the group ones are drawn from
the same parent population is 10 per cent, and that for the group and the field ones
is 35 per cent.
We also evaluate the photometric similarity using the Mann-Whitney $U$ test.
The magnitude distributions of cluster and field red galaxies
are not statistically different at 15 per cent level.
The probability for no difference between cluster and group red
galaxies is 35 per cent, and that for group and field ones is 5 per cent.
The probabilities that the colour distributions
are the same are 50 per cent for all the environments.
Therefore, the photometric properties are similar in most environments.
They might differ in some environments, but the possible differences
are small and they are unlikely to affect our conclusions significantly.

Fig. \ref{fig:composite_spectra} compares the composite spectra.
The continua of all the spectra are very similar.
They all show a strong 4000$\rm\AA$ break, which is typical for evolved red galaxies.
They do not actively form stars as indicated by their  \oii\  emission.
Even the field red galaxies have only a weak \oii\ emission, EW\oii$=4\rm\AA$.
The fractions of \oii\ emitters are 10, 10, and 30 per cent in cluster, group, and field.
It is interesting to note that the strengths of the \oii\ emissions are
weaker than those for $z\sim0.8$ red galaxies.
\citet{tanaka06} showed that, at $z\sim0.8$, cluster red galaxies do not show
any \oii\ emission, while group and field red galaxies have EW\oii$=4\rm\AA$
and 13$\rm\AA$ on average, respectively.
More active star formation is seen in less dense environments at higher redshifts.

\subsection{Spectral Diagnostics}
\label{sec:spectral_diagnostics}

We now quantify the differences between the composite spectra.
We measure the strengths of the $4000\rm\AA$ break ($D_{4000}$) and H$\delta$ absorption
since they are sensitive to star  formation histories on different time scales.
Details of the measurement scheme and error analysis are described in \citet{tanaka06}.
Note that the correction for H$\delta$ emissions from gaseous nebulae
for galaxies at $z>0$ is revised.
To correct for the H$\delta$ emission filling, we derive
a correlation between the amount of H$\delta$ emission (which is estimated
from H$\alpha$ emissions assuming the 'case B' recombination; \citealt{osterbrock88})
and EW\oii\ using red galaxies at $z=0$.
The derived correction values are about half as small as those adopted in \citet{tanaka06}.
There is a large scatter in the correlation between H$\delta$ emission and EW\oii,
but we have confirmed that the uncertainties arising from the correction do not alter our conclusions.

In addition to the $z\sim0.5$ and 0.8 samples, we construct a local sample for comparison
from the SDSS \citep{york00}.
We extract galaxies from the fourth public data release \citep{adelman06}.
We select red galaxies on the CMR in a similar way to $z\sim0.8$ and $z\sim0.5$ galaxies.
We do not separate $z=0$ galaxies into the cluster, group, and field environments
since the differences between the environments are very small compared to
the differences observed at $z>0$.
Note that we use $M_V\lesssim M_V^*+1$ galaxies at all redshifts.
In what follows,  we often refer to our previous papers.
We compile results from \citet{tanaka05,tanaka06} and this work and
they are summarized in Table \ref{tab:results_summary}.

\begin{figure}
\begin{center}
\leavevmode
\epsfxsize 1.0\hsize \epsfbox{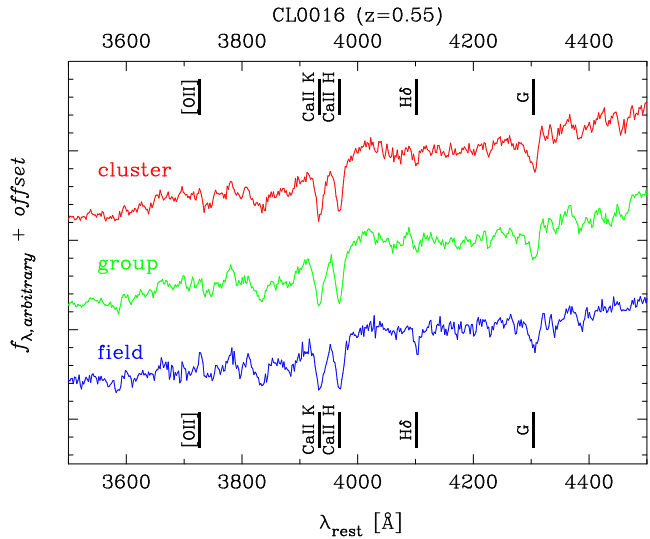}
\end{center}
\caption{
The rest-frame composite spectra of cluster, group and field galaxies.
}
\label{fig:composite_spectra}
\end{figure}

\begin{figure*}
\begin{center}
\leavevmode
\epsfxsize 0.48\hsize \epsfbox{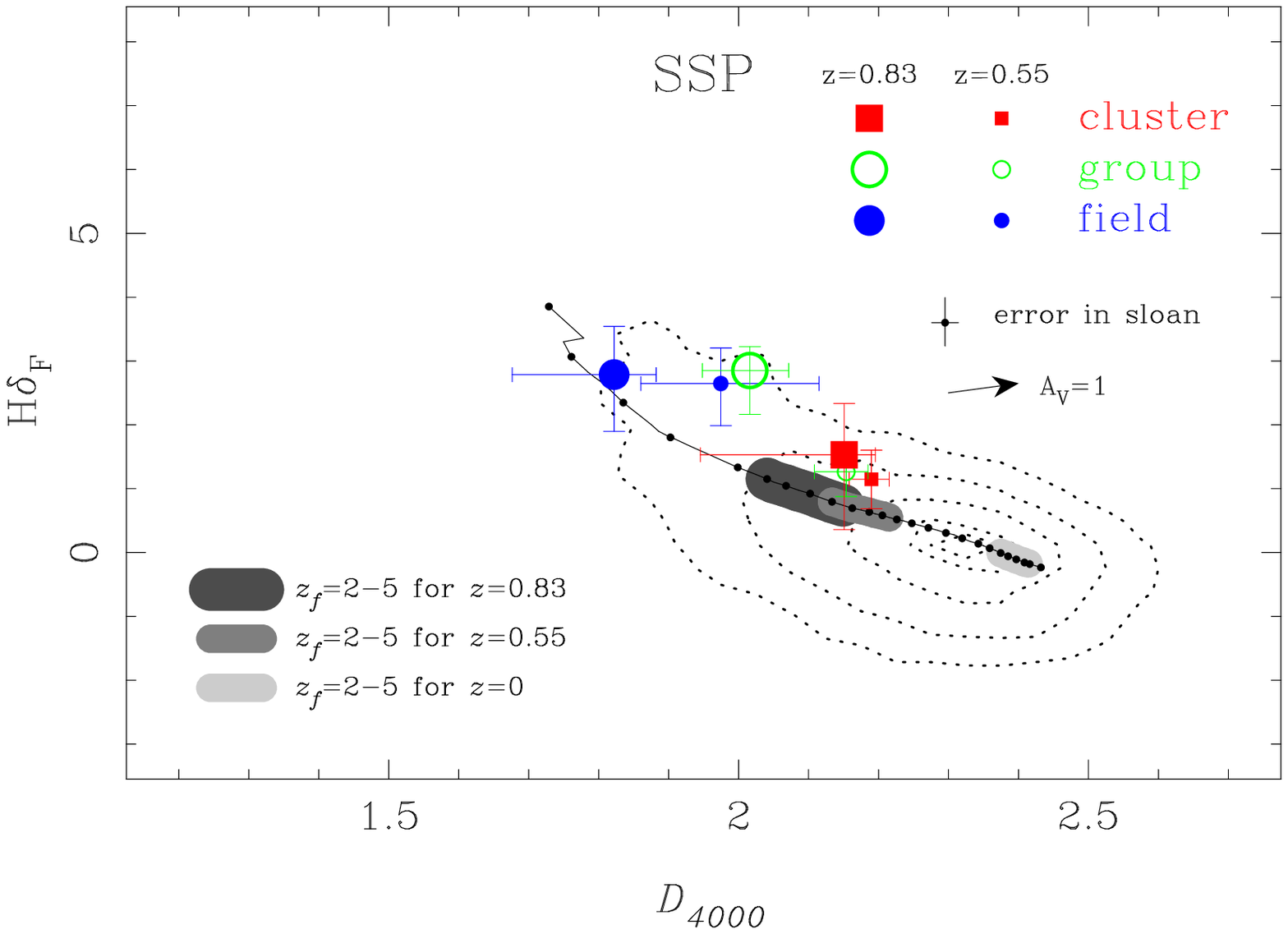}\hspace{0.5cm}
\epsfxsize 0.48\hsize \epsfbox{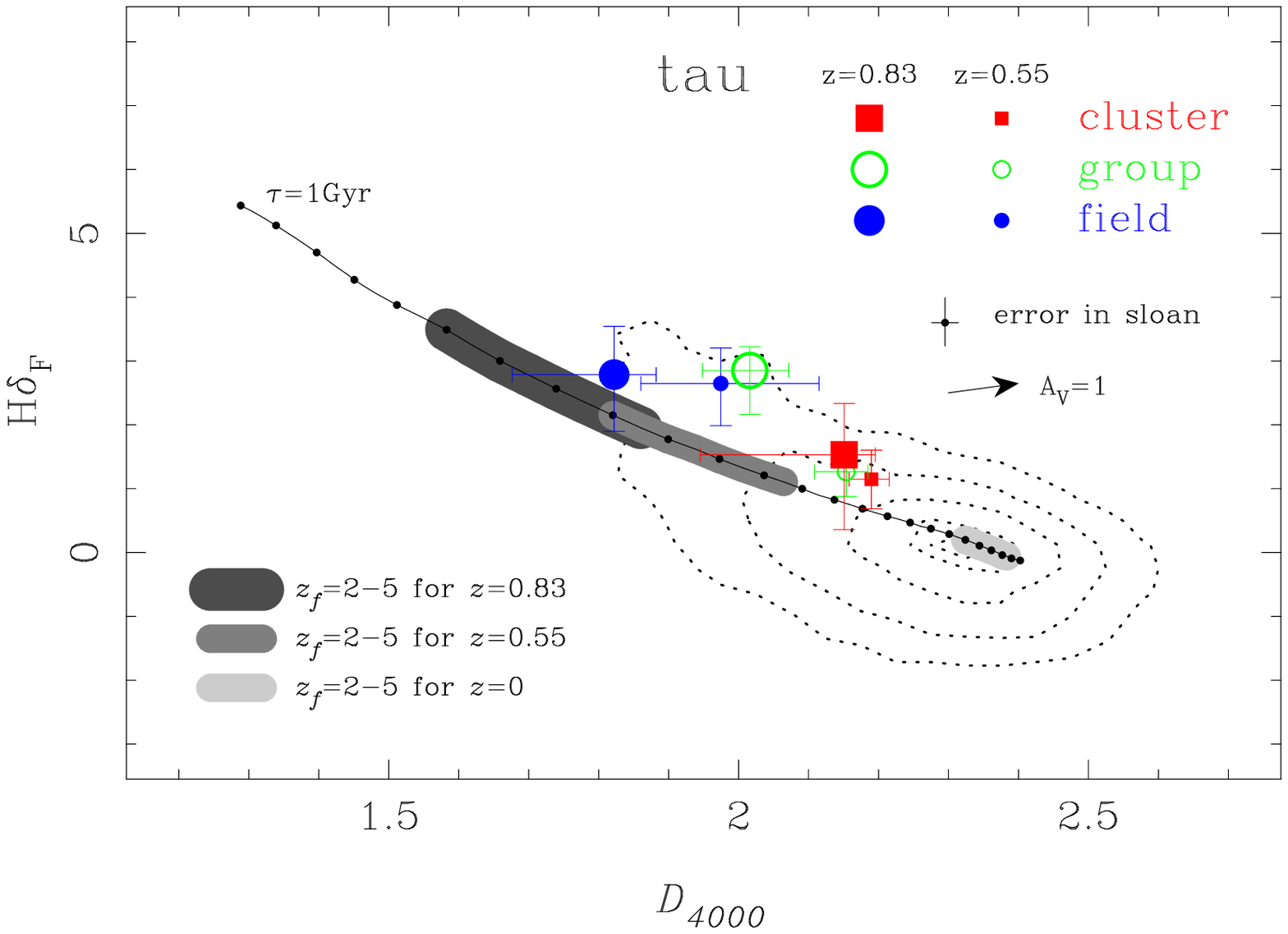}\\\vspace{0.5cm}
\epsfxsize 0.48\hsize \epsfbox{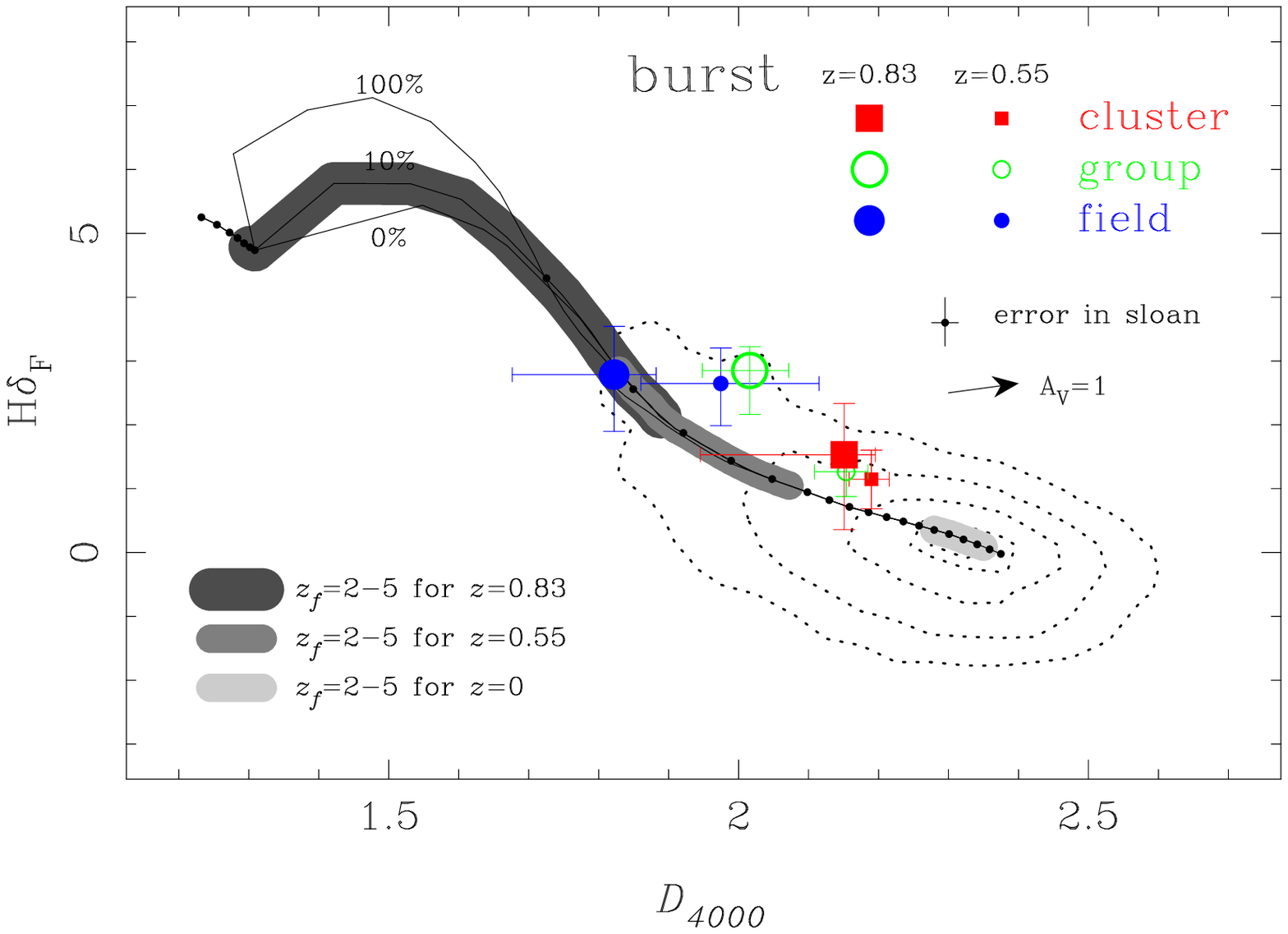}
\end{center}
\caption{
The H$\delta_F$ index plotted against $D_{4000}$. The filled square, open
circle and filled circle, respectively, show cluster, group and field composites
as shown in each panel.
The contours show distribution of red galaxies at $z=0$
brighter than $M^*_V + 1$, and enclose 5, 20, 50, 80 and 95 per cent of the galaxies.
A typical measurement error for the $z\sim0$ galaxies and the $A_V=1$ mag. vector
are indicated in each panel.
The panels show different BC03 model tracks; SSP, tau and burst models
from top to bottom. The model predictions are presented as the line-connected
points (points at every 0.5 Gyr). The model starts from 1 Gyr and ends at
13 Gyr (from left-hand side to right-hand side). The regions shaded dark grey,
grey, and light grey, respectively, show the model locus of $z_f =2$ to 5 for galaxies
at $z = 0.83$, 0.55 and 0, where $z_f$ is the formation redshift of model galaxies. Note
that, in the burst model, $z_f = 2$ and 5 for $z = 0.83$ galaxies correspond to
1 Gyr before and 1 Gyr after the burst.
For $z=0.55$ galaxies, they correspond to 1 and 3 Gyr after the burst.
For $z = 0$, 6 and 8 Gyr after the burst.
}
\label{fig:spec_diag}
\end{figure*}

We present in Fig. \ref{fig:spec_diag} the distribution of $D_{4000}$ and H$\delta_F$
indices of red galaxies at $z \sim 0.8$ (large symbols), $z\sim0.5$ (small symbols) and $z = 0$ (contours)
along with model predictions.
Here we present the same models adopted in \citet{tanaka06},
which make use of the \citet{bruzual03} population synthesis model (BC03 model hereafter).
As a default parameter set, we adopt the Chabrier initial mass function between
$0.1-100\rm M_\odot$, solar metallicity, and no dust extinction
(see \citealt{bruzual03} and references therein).
Since we discuss bright galaxies, the assumption of solar metallicity is reasonable.
Also, since the red galaxies are dominated by red old stars
(see Fig. \ref{fig:composite_spectra}), we assume no dust extinction to start with.
We will come back to the effects of changing metallicity and extinction later.
Three star formation histories are employed in the model: single burst, exponential decay with
a time scale of $\tau=1$ Gyr, and burst + sharp truncation\footnote{
In this model, the star formation rate is constant for the first 4 Gyr and then
star formation is sharply truncated with a burst.
Three burst strengths are adopted; stars newly born in the burst amount to
0, 10, 100 per cent of the existing stars.
The 0 per cent burst means that star formation is truncated without a burst.
No star formation occurs after the truncation.
} models.
We refer to these histories as SSP, tau, and burst models.
As shown in Fig. \ref{fig:spec_diag},
our models reasonably cover the distribution of the observed $z = 0$ galaxies.
This ensures that the models describe typical star formation histories of galaxies.
Here we separately discuss the cluster, group, and field environments for clarity.

\noindent
{\bf Cluster :}
The cluster red galaxies at $z\sim0.8$ and 0.5 are consistent
with passive evolution within the error.
They are likely to evolve to the normal red galaxies at $z=0$
($D_{4000}\sim2.3$ and H$\delta_F\sim0$).

\noindent
{\bf Group :}
In contrast to the cluster galaxies, the group  red galaxies
at $z\sim0.8$ cannot be reproduced by the simple evolution models
such as SSP and tau.
Their $D_{4000}$ is consistent with passive evolution,
but their H$\delta$ absorptions are too strong.
By $z\sim0.5$, H$\delta$ absorptions of group red galaxies are weakened and
they get on passive evolution.
Their star formation is quenched (the \oii\ emission disappears), and they will
passively evolve to red galaxies down to $z=0$.

\noindent
{\bf Field :}
Field red galaxies at $z\sim0.8$ can be reproduced by the tau and the burst models.
They still show a sign of star formation as indicated by
\oii\ emissions (EW\oii$=13\rm\AA$; Table \ref{tab:results_summary}).
Their star formation is not completely quenched yet despite their red colours.
Star formation activities of field red galaxies weakens from $z\sim0.8$
to $z\sim0.5$ (EW\oii$=4\rm\AA$).
Field red galaxies at $z\sim0.5$ have similar spectral properties to those of group
galaxies at $z\sim0.8$
and only the burst model gives an acceptable fit.
Their H$\delta$ absorption is strong for their $D_{4000}$.

Galaxies with a large $D_{4000}$ and a strong H$\delta$ absorption
(i.e. group galaxies at $z\sim0.8$ and field galaxies at $z\sim0.5$)
are very interesting.
Their $D_{4000}$ are consistent with passive evolution within the errors
(field galaxies at $z\sim0.5$ are marginally consistent),
but H$\delta$ absorptions are too strong.
In fact, we fail to reproduce group red galaxies at $z\sim0.8$
with any of our models.
We focus on these galaxies in what follows.
As shown later, these galaxies have important implications
for the truncation of star formation.
We note that the correction applied for the emission filling is small
($\Delta$H$\delta_F=+0.2$) and thus the strong H$\delta$ absorption
is not caused by the error in the emission correction.
Note as well that the composite spectra at $z=0.55$ and 0.83 are made from
$\sim10$ galaxies each and hence their observed offsets in the $D_{4000}$
and H$\delta_F$ diagram is underestimated when compared directly
to those of the typical $z=0$ galaxies.
In fact, if we make a composite spectrum of 10 galaxies for $z=0$ as well,
the observed scatter in H$\delta_F$ decreases by a factor of
$\sim\sqrt{10}$, and the fraction of galaxies with H$\delta_F>2$
around $D_{4000}\sim2$ at $z=0$ is only 2 per cent.

We further explore the parameter space of the models.
We have two adjustable parameters left, namely, metallicity and dust extinction.
On one hand, $D_{4000}$ increases with increasing dust extinction
(adding extinction makes galaxies redder, i.e. larger $D_{4000}$).
Also, it is sensitive to metallicity (increasing metallicity makes galaxies redder).
On the other hand, an H$\delta$ absorption is not sensitive to both of them.
The H$\delta_F$ index is measured in a narrow wavelength window and it is
almost insensitive to dust extinction.
It does not strongly change with metallicity either since it is a Hydrogen line\footnote{
If we adopt super-solar and sub-solar metallicity models ($Z=0.05$ and 0.008),
the difference in H$\delta_F$ from the solar metallicity model
($Z=0.02$) is $\sim0.3\rm\AA$ at most.}.
Thus, dust extinction and metallicity can easily change $D_{4000}$
while keeping H$\delta$ nearly unchanged.
The opposite (i.e. changing H$\delta$ keeping $D_{4000}$ unchanged)
is very difficult to achieve due to the nature of the indices.
An important point here is that only the recent star formation history
can change the behaviour of H$\delta$
(see Fig. \ref{fig:spec_diag}).

Keep this in mind, let us go back to Fig. \ref{fig:spec_diag}.
As the tau model demonstrates, a gradual truncation of star formation
does not enhance the H$\delta$ absorption and fail to reproduce the observed strong H$\delta$ absorption.
The strong H$\delta$ absorption then favours a scenario that
the star formation is suppressed on a short time-scale rather than a slow decline.
But, still, the burst model does not fit the red galaxies in groups at $z\sim0.8$.
We find that an $A_V\sim1$ mag. extinction\footnote{
Here we assume that the extinction affects stars of all ages equally
(i.e. we ignore extinction within H{\sc ii} regions).}
gives the burst model an acceptable fit
to group galaxies at $z\sim0.8$ and field galaxies at $z\sim0.5$.
This level of extinction is reasonably expected in a starburst population \citep{obric06}.
Of course, we could use the tau model with extinction to fit the observation.
However, the tau model requires $A_V\sim2$ mag. to fit $z\sim0.8$ group galaxies,
which is apparently too large for a gradual truncation model.
The composite spectrum shows the strong 4000$\rm\AA$ break and Ca{\sc ii}HK absorptions,
and the galaxies are apparently dominated by red old stars
(see Fig. 7 of \citealt{tanaka06}).
Thus, it is unlikely that they have a large amount of dust as $A_V\sim2$.
Therefore, the most natural interpretation is that their strong H$\delta$
absorption is caused by a sharp decline in their star formation rates,
and their large $D_{4000}$ reflects a large fraction of evolved red stars
with a reasonable amount of dust.

To sum up, the strong H$\delta$ absorption observed in the $z\sim0.8$ group
and the $z\sim0.5$ field galaxies suggests that their star formation activities
have been truncated on a short time scale.
In the next section, we discuss possible physical processes that triggered the suppression
of star formation activities.

\begin{table*}
\caption{
Summary of results from \citet{tanaka05,tanaka06} and this work.
Status of the CMR in the field, group, and cluster environments along with
the strengths of \oii\ emissions of red galaxies are shown.
}
\label{tab:results_summary}
\begin{tabular}{lccc}
\hline
Redshift & Field  & Group & Cluster\\
\hline
$z\sim0.8$ & No clear CMR      & CMR at bright end & tight CMR\\
           & EW\oii$=13\rm\AA$ & EW\oii$=4\rm\AA$  & no \oii\\
\hline
$z\sim0.5$ & CMR at bright end & tight CMR         & tight CMR\\
           & EW\oii$=4\rm\AA$  & no \oii           & no \oii\\
\hline
\end{tabular}
\end{table*}

%
%
\section{Discussion}
\label{sec:discussion}

\subsection{Decline in Star Formation Rates and Build-up of CMR}

Let us jointly discuss implications of our results from
\citet{tanaka05,tanaka06} and this work for galaxy evolution.
We observed the build-up of the CMR since $z\sim0.8$.
The CMR first appears at the bright end and the faint end appears later.
Interestingly, the build-up is 'delayed' in lower density environments
(but see also \citealt{delucia06}).
The CMR build-up and the evolution in \oii\ emissions are summarized
in Table \ref{tab:results_summary}.
In what follows, we discuss physical drivers of the truncation of
star formation and build-up of the CMR using Fig. \ref{fig:spec_diag}
and Table \ref{tab:results_summary}.

We spectroscopically confirm that the truncation of star formation is
delayed in lower-density environments.
While the colour-magnitude distributions of the cluster and the group red
galaxies are similar, the composite spectrum of the group red galaxies
at $z\sim0.8$ show a weak \oii\ emission with a strong H$\delta$ absorption,
which is not seen in the cluster red galaxies (Fig. 7 of \citealt{tanaka06}).
This means that the group red galaxies at $z\sim0.8$ must have had recent
star formation activities, while the cluster red galaxies 
at $z\sim0.8$ have not formed stars for a long time.
Group galaxies have stopped their star formation by $z\sim0.5$
(the \oii\ emission disappears)
and they get on passive evolution as indicated in Fig. \ref{fig:spec_diag}.
The field galaxies will stop forming stars even later since the field
red galaxies at $z\sim0.8$ are probably still  forming stars (EW\oii$=13\rm\AA$),
and there is still a sign of weak star formation at $z\sim0.5$ (EW\oii$=4\rm\AA$).
It is reasonable to consider that cluster red galaxies at $z\sim0.8$ have
stopped star formation well in advance, while group red galaxies at
$z\sim0.8$ and field red galaxies at $z\sim0.5$ are just in the process of truncation.

The truncation of star formation activities is directly mirrored to the build-up of CMR.
The fact that the red cluster galaxies at $z\sim0.8$ show no star formation activities
is consistent with our earlier results from the photometric data that
the CMR is already built up down to faint magnitudes \citep{tanaka05}.
Group galaxies at $z\sim0.8$ form a CMR at the bright end, but not at the faint end.
This should mean that the bright end of the relation is built up shortly before $z\sim0.8$.
That is, the star formation rates of bright galaxies in groups dropped in a recent past.
In fact, the composite spectrum shows only a small amount of residual star formation
as indicated by the weak \oii\ emission (we probe only bright galaxies with spectroscopy).
The \oii\ emission disappears and the residual star formation is completely
quenched by $z\sim0.5$ and the CMR extends fainter magnitudes.
Field red galaxies at $z\sim0.8$ show a sign of star formation activity 
and they do not form a tight CMR.
The bright end of the field CMR appears at $z\sim0.5$.
The composite spectra reflect this --- the \oii\ emission significantly weakens
from $z\sim0.8$ to $z\sim0.5$.

What physical mechanism truncates star formation activities?
A hint for this question should lie in the groups at $z\sim0.8$ and in the
field at $z\sim0.5$ since they are likely in the process of truncation.
A close inspection of them will therefore give us
a clue to the physical driver of the truncation.

\subsection{Physical Process}

Intensive studies on relationships between galaxy properties and
environment in the local Universe have shown that red early-type galaxies
dominate high-density environments, and blue late-type galaxies preferentially
live in low-density environments \citep{dressler80,balogh99,lewis02,gomez03,tanaka04,blanton05}.
Several physical mechanisms are claimed
to drive the observed environmental dependence,
e.g. ram-pressure stripping, \citep{gunn72}, and strangulation \citep{larson80}.
Each mechanism has its specific environments in which it works most effectively.
For example, ram-pressure stripping is most effective in the cores of
rich clusters and galaxy-galaxy interaction is most effective in groups.
Thus, observations of galaxies in various environments put constraints
on the processes at work.

Here we focus on poor groups of galaxies.
Groups have been paid less attention than custers, but
observations of poor groups have shown that the fraction of red early-type
galaxies is larger in groups than in the field.
For example, \citet{zabludoff98} and \citet{tran01} studied nearby poor groups
and found that properties of galaxies in the poor groups differ
from those in the field, in the sense that a larger fraction of
red and early-type galaxies reside in groups.
Based on a large sample of galaxies delivered by the 2dF survey,
\citet{martinez02} showed that the fraction of non-star-forming galaxies
is higher in groups as poor as $\rm 10^{13} M_\odot$ compared
with that of the field.
\citet{weinmann06} showed that the fraction of early-type galaxies
increases with increasing mass of clusters.
\citet{kodama01} and \citet{tanaka05} found that poor groups surrounding
rich clusters are dominated by red galaxies.
It is thus likely that galaxies start to change their properties
in groups before they finally merge into rich clusters.

What do these observations tell us?
They actually put strong constraints on the proposed mechanisms.
If we have only the cluster-specific mechanisms, we fail to
reproduce the observation.  Because no mechanisms work on
galaxies in poor groups, and thus we expect that the statistical
properties of the group galaxies will be similar to those of the
field galaxies.
This is inconsistent with the observation in which we see a
larger fraction of red early-type galaxies in groups than in the field.
Although cluster-specific mechanisms work on some galaxies
\citep{kenney04,vollmer04}, this suggests that we need an alternative mechanism
as a primary driver of the truncation of star formation activities in
low density environments.

We are left with two viable mechanisms: galaxy-galaxy interactions
(e.g. \citealt{mihos96}) and strangulation \citep{larson80,balogh00}.
Both are effective in groups.
Galaxy-galaxy interactions trigger starbursts and star formation is
truncated after the burst. Strangulation gradually truncates star formation over
$\sim1$ Gyr. Thus, we can test these processes with the H$\delta$ absorption
-- starburst plus truncation enhances the H$\delta$ absorption after
the burst, while strangulation does not trigger the enhancement
as shown in Fig. \ref{fig:spec_diag}
(strangulation should follow a similar track to the tau model).

We observe strong  H$\delta$ absorptions in groups at $\sim0.8$
and field at $z\sim0.5$.
We recall that these are the environments in which we observed
the on-going build-up of the CMR, and thus galaxies in these
environments are likely in the process of
the truncation of star formation.
As we have discussed above, the strong H$\delta$ absorption
suggests that star formation is quenched on a short time scale.
The sharp decline in their star formation rates then favours
the interaction scenario over strangulation.
Galaxies may interact with one another in these environments,
their star formation rates drop sharply, they become red
and form a tight CMR.
The fact that field red galaxies at $z\sim0.5$ also show a strong H$\delta$
absorption may lend further support to the interaction scenario since
strangulation is not effective in the field\footnote{
Galaxy-galaxy interactions are also less effective
in the field compared to groups due to the low galaxy density.
The similarity in the composite spectra suggests that 
the interaction rate in the $z\sim0.5$ field is comparable to that in
the $z\sim0.8$ groups, contrary to our expectations.
It should be noted, however, that the $z\sim0.5$ field spectrum
is made from only 7 galaxies and an increased sample is needed to
confirm the picture we discuss here.
}.
Taking all these circumstantial evidence, we suggest that
a galaxy-galaxy interaction is the driving process of the truncation of star formation
and the build-up of the CMR.

But, would galaxy-galaxy interactions drive the down-sizing?
Mergers may not selectively occur between massive galaxies,
and we may fail to explain down-sizing with mergers only.
A possible process responsible for the down-sizing phenomenon
recently emerged from galaxy formation models.
Recent semi-analytic models suggest that energy feedback from active galactic nuclei (AGN)
is a key ingredient to reproducing the down-sizing phenomenon
(e.g. \citealt{bower06}; \citealt{croton06}).
Galaxies grow hierarchically, and the central black holes grow with time.
The energy feedback from the AGN activity is stronger
for more massive black holes (i.e. for more massive halos).
This results in a selective suppression of gas cooling
in massive halos, i.e. down-sizing.
Once the cold gas in a massive object is exhausted, star formation will not
take place any more and the object in the halos remain red and stay on the CMR afterwards.
For less massive galaxies, the AGN feedback is not strong enough and
gas continues to cool.
The down-sizing might be caused by two processes:
the truncation of star formation and the suppression of gas cooling.
That is, we need (1) a process that makes blue galaxies red and
(2) a process that keeps them red.
The former would be interactions as we have discussed above
and the latter would be the AGN activity.

In this paper, we rely on the spectroscopic information to suggest the interaction scenario.
A direct way to prove it is to see morphology of galaxies.
If, for example, a fraction of interacting galaxies in groups at $z\sim0.8$
is large compared to the other environments, it will be strong evidence for
the interaction scenario.
The resolution of our Subaru images is good ($\sim0.6$ arcsec) for ground-based
observations, but it is not good enough to study morphologies at high redshifts
($0.6$ arcsec corresponds to 4.6 kpc at $z\sim0.8$ in a physical scale).
It is therefore essential to obtain deep high-resolution ACS/HST images.
We will report on galaxy morphology in $z\sim0.8$ groups in a forthcoming paper
(Demarco et al. in prep).

%
%
\section{Summary and Conclusions}
\label{sec:summary}

We have carried out spectroscopic observations of the photometrically
identified large-scale structure around the CL0016 cluster at $z=0.55$.
We spectroscopically confirm a huge filament that goes
in the N-S direction connecting the clumps extended over $20h_{70}^{-1}$Mpc
and a filament extending westward from the main CL0016 cluster.
Several clumps are embedded in the filaments, and they are likely
bound to the main CL0016 cluster.
This is one of the most prominent large scale structures ever found in the Universe.

We make composite spectra of field, group, and cluster red galaxies.
They show a strong $4000\rm\AA$ break, which is typical for evolved red galaxies.
Group and cluster red galaxies show no sign of star formation, while
field red galaxies show a small amount of residual star formation as
indicated by the weak \oii\ emission.
By combining the composite spectra at $z\sim0.8$ from \citet{tanaka06} with those at $z\sim0.5$,
we spectroscopically confirm the environmental dependence of star formation activities ---
more active star formation is seen in less dense environments at higher redshifts, and vice versa.
This trend is closely mirrored to the build-up of the CMR.

We then quantify the strengths of the H$\delta$ absorption and $4000\rm\AA$ break.
Red galaxies in groups at $z\sim0.8$ and field at $z\sim0.5$
show a strong H$\delta$ absorption for their $D_{4000}$.
Interestingly, these are the environments in which we observed the on-going
build-up of the CMR \citep{tanaka05}.
Therefore, galaxies tend to show a strong H$\delta$ absorption
when they stop forming stars.
There are several mechanisms claimed to affect galaxy properties.
Recent observations suggest that we need a process effective
in low density regions such as groups.
We have two viable mechanisms for this: galaxy-galaxy
interactions and strangulation.
The observed strong H$\delta$ absorption favours the scenario that
star formation is truncated on a short time scale.
Therefore, we suggest that
galaxy-galaxy interactions are likely the primary process behind the
truncation of star formation and hence responsible for
the build-up of the CMR.

%
%
\section*{Acknowledgements}
We thank the anonymous referee for his/her suggestions, which improved the paper.
M.T. acknowledges support from the Japan Society for Promotion of Science (JSPS)
through JSPS research fellowships for Young Scientists.
This work was financially supported in part by a Grant-in-Aid for the
Scientific Research (No.\, 15740126, 18684004) by the Japanese Ministry of Education,
Culture, Sports and Science.
This study is based on data collected at Subaru Telescope, which is operated by
the National Astronomical Observatory of Japan. 

Funding for the creation and distribution of the SDSS Archive has been
provided by the Alfred P. Sloan Foundation, the Participating Institutions,
the National Aeronautics and Space Administration, the National Science Foundation,
the U.S. Department of Energy, the Japanese Monbukagakusho, and the Max Planck Society.
The SDSS Web site is http://www.sdss.org/.

%
%

\end{document}